\documentclass[lettersize,journal]{IEEEtran}
\usepackage[latin1]{inputenc}
\usepackage{times,amsmath}
\usepackage{amssymb}
\usepackage{pstool}
\usepackage{multirow,multicol}
\usepackage{enumerate}
\usepackage{graphicx}
\usepackage{MnSymbol}
\usepackage{stfloats} 
\usepackage[table]{xcolor}
\usepackage[square, comma, sort&compress, numbers]{natbib}
\usepackage{epstopdf}
\AppendGraphicsExtensions{.tif}
\usepackage{hyperref}
\usepackage{algorithm}
\usepackage{algpseudocode}
\usepackage{subfig}
\graphicspath{ {Figures/} }
\usepackage[export]{adjustbox}

\begin{document}


\title{Malicious Pseudo-Ranging and Localization of Static LOS Wireless Users via Downlink Modulation Classification and Uplink Refinement}



\author{Ali Hanif, \IEEEmembership{Student Member, IEEE}, Abdulrahman Katranji, Nour Kouzayha, \IEEEmembership{Member, IEEE},\\ Muhammad Mahboob Ur Rahman, \IEEEmembership{Senior Member, IEEE}, and Tareq Y. Al-Naffouri, \IEEEmembership{Fellow, IEEE}%
\thanks{The authors are with the Department of Electrical and Computer Engineering, CEMSE Division, King Abdullah University of Science and Technology  (KAUST), Saudi Arabia.}
\vspace{-0.5cm}}

\maketitle

\begin{abstract} 


The broadcast nature of the wireless medium and openness of wireless standards, e.g., 3GPP releases 16-20, invite adversaries to launch various active and passive attacks on cellular and other wireless networks. This work identifies one such loose end of wireless standards and presents a novel passive attack method enabling an eavesdropper (Eve) to localize a line-of-sight stationary wireless user (Bob) who is communicating with a base station or WiFi access point (Alice). The proposed attack involves two phases. In the first phase, Eve performs modulation classification by intercepting the downlink channel between Alice and Bob. This enables Eve to utilize the publicly available modulation and coding scheme tables to do pesudo-ranging, i.e., the Eve determines the ring within which Bob is located, which drastically reduces the search space. In the second phase, Eve sniffs the uplink channel, and employs multiple strategies to further refine Bob's location within the ring. In simulations, the proposed attack is validated for single-user, multi-user, and multiple-antenna scenarios. Towards the end, we present our thoughts on how this attack can be extended to other scenarios such as non-line-of-sight and mobile users e.g., cars, drones, pedestrians, and how this attack could act as a scaffolding to construct a malicious digital twin map. 
\end{abstract}

\begin{keywords}
Modulation and coding scheme, modulation classification, physical layer security, passive attack, eavesdropping, ranging, and localization.
\end{keywords}

\section{Introduction}

\IEEEPARstart{T}{he} open nature of wireless medium invites adversaries to launch an ever-increasing spectrum of active and passive attacks, i.e., it is always possible to intercept, jam, and manipulate the ongoing communication between the legitimate nodes of a cellular/WiFi network through a low-cost, off-the-shelf spectrum analyzer or a software-defined radio~\cite{chorti2022context}. Such attacks, when occur, lead to a wide range of problems, e.g., false data injection, loss of data integrity, breach of data confidentiality, service outage, etc., and therefore, pose a great threat to 6G communication systems in particular~\cite{mucchi2021physical}, and all kinds of wireless communication networks at large~\cite{aman2023security}. Recently, there have been some reactive attempts by key stakeholders from the industry to design additional countermeasures to thwart a subset of adversarial attacks on cellular networks after they had actually occurred. For example, the 3rd generation partnership project (3GPP) has published a technical report (TR 33.809, Release 16) that provides a list of 5G security enhancements against false base stations (also known as rogue or fake base stations) that impersonate legitimate network elements in order to intercept, manipulate, or degrade communications~\cite{3gppTR33.809}. Nevertheless, in the era of generative artificial intelligence (AI), when there is a considerable mind shift to make the individual components of future 6G networks more transparent and AI-native through novel concepts such as open radio access network (RAN) systems~\cite{polese2023understanding}, software-defined networking (SDN), it is high time to reassess the security profile of future 6G systems.


Moreover, the openness of various wireless standards may allow attackers to identify additional vulnerabilities in cellular/WiFi networks and capitalize on them. In fact, this work discovers one such loose end and, thereafter, presents a novel passive attack from the ethical hacking perspective. Specifically, this work focuses on modulation and coding scheme (MCS) tables that are routinely published by the 3GPP under releases 1x (\cite{3gppTS38.213}, see TS 38.213/38.214 specification by 3GPP) and by IEEE under releases 802.11xx~\cite{9442429}. Traditionally, MCS tables are utilized by wireless networks to realize adaptive modulation and coding (AMC), which is a mandatory operation that aims to make the best use of the fading wireless channel~\cite{4224371}. In order to help the base station (BS) utilize the MCS table, the user equipment (UE) measures the channel quality on the downlink and computes the channel quality index (CQI), which is then mapped to an MCS index. Accordingly, a modulation scheme and a coding rate are picked by the BS for the downlink communication during the next slot. In short, MCS-based AMC is an attempt to realize intelligent communication. Nevertheless, the fact that MCS tables are public information could expose the cellular/WiFi networks to attackers, as explained in the rest of this paper.

This work studies an innovative scenario whereby a malicious sensing node (Eve) sits close to a BS and intercepts the ongoing communication between the BS and a number of UEs, on both downlink and uplink. Essentially, this is the same as the old-school concept of cognitive radio~\cite{mitola1999cognitive}, except that Eve now aims to do location sensing via the MCS tables, instead of traditional spectrum sensing.

{\bf Contributions:} 
This work introduces a novel passive attack that allows a malicious eavesdropper to obtain the approximate range and location estimates of one or more line-of-sight (LOS) stationary users associated with a wireless cellular base station or a WiFi access point. 
The main contributions of this work are as follows:
 \begin{enumerate}
    \item {\it Pseudo ranging:} Eve sniffs the broadcast transmission of the base station intended for licensed user(s) during the downlink phase, in order to perform modulation classification on the intercepted signal(s). Thereafter, Eve utilizes the publicly available MCS tables in order to reverse-map the detected modulation scheme to an MCS index/CQI, to obtain upper and lower bounds on the range (i.e., the distance between the user(s) and the base station). This significantly narrows down the search space--from a full cell coverage region down to a ring.
    \item {\it UE Localization:} Eve sniffs the transmission of the user during the uplink phase, while traversing through the ring (i.e., the narrowed search space) where a given user is likely located. The sniffing at many grid points during the circular motion of Eve inside the ring allows Eve to record signal-to-noise ratio (SNR) at each grid point and declare its location where the SNR is maximum as the initial location estimate of Bob.
    Eve further does a coarse ranging followed by another circular motion in order to obtain a precise location estimate of the user.
 \end{enumerate}

{One core objective of this work is to raise awareness about vulnerabilities of modern cellular networks due to the open nature of wireless cellular standards documents, through ethical disclosure of one practical attack method.}

This work makes a fundamental contribution to the field of physical layer security as it introduces a novel passive attack, first of its kind~\cite{mitev2023physical}.  
Further, this work stands out compared to previous works on non-cooperative source localization due to the fact that it is based on passive sensing only and doesn't require the deployment of any anchor nodes with known locations~\cite{8734111}. Finally, this work, being a malicious sensing method, also contributes to integrated sensing and communication (ISAC) methods of future 6G systems~\cite{kaushik2024toward}.


\subsection{Potential impact}
This work demonstrates that by reverse-engineering MCS tables, an adversary can map observed MCS indices to SNR ranges and corresponding geographic regions, enabling unsolicited pseudo-ranging and localization in a covert manner. The implications of proposed covert attack are grave as it threatens spatial privacy, allowing long-term tracking of wireless cellular users, malicious reconstruction of mobility maps, and deanonymization, while also serving as a precursor to targeted jamming, spoofing, with potentially severe consequences in critical systems such as vehicular or drone networks. Its stealthy nature makes detection difficult, underscoring both the urgency of reporting such first-of-its-kind attacks and the broader need to re-examine public wireless standards for exploitable vulnerabilities with serious implications.

\begin{figure*}
    \centering
    \includegraphics[width=0.99\textwidth,keepaspectratio]{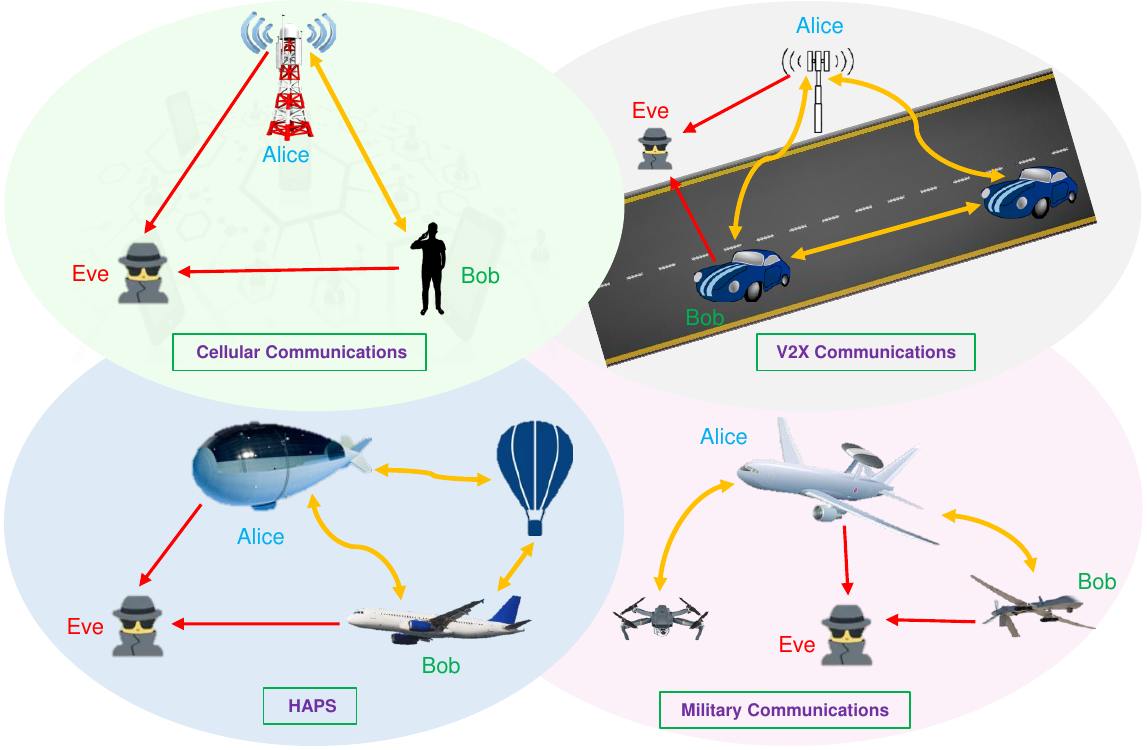}
    \caption{Potential application scenarios of the proposed modulation classification-based passive attack.}
    \label{scenario}
\end{figure*}

The proposed infrastructure-less passive attack can be employed to localize various types of users as shown in Fig.~\ref{scenario}, including cellular users, users in vehicle-to-everything (V2X) networks, users communicating with non-terrestrial nodes, e.g., high altitude platform systems (HAPS), and military unmanned aerial vehicles (UAV)/drones connected to airborne warning and control systems (AWACS).

\color{black}
\section{System Model} \label{sys_model}
We consider a system model comprising a {BS}/access point (Alice), one UE node (Bob), and one adversary node (Eve), which has the malicious intention to localize Bob. A ray tracing channel model with additive white Gaussian noise (AWGN) is used to model all three communication channels, i.e., the pair-wise channels between Alice, Bob and Eve. For simplicity, we assume that all the nodes are equipped with single antennas. We consider a LOS scenario in order to assess the performance of the proposed modulation classification-based attack in estimating the location of the UE. Alice initiates the downlink communication by receiving the channel quality indicator from Bob and mapping it to a modulation and coding scheme using a standard MCS table shown in Table~\ref{MCStable}. 
Multiple frequency bands ($5$~GHz, $28$~GHz, and $100$~GHz) and various transmit powers of Alice are considered to validate the proposed approach under different scenarios. Finally, the scenarios involving two UE nodes, and multiple antennas at the adversary node Eve are also discussed. 

\begin{table}[h!]
    \centering
    \caption{MCS table for IEEE 802.11ac wireless local area network (WLAN). VHT stands for very high throughput. IEEE 802.11ac offers 20 MHz bandwidth, and two frame intervals of 800 ns and 400 ns.  }
    \begin{tabular}{|c|c|c|c|c|c|}
    \hline
         \multirow[c]3{*}[0in]{\shortstack{\textbf{VHT}\\\textbf{MCS}}} & \multirow[c]{3}{*}[0in]{\textbf{Modulation}}  & \multirow[c]{3}{*}[0in]{\textbf{Coding}}   & \multicolumn{3}{|c|}{\textbf{20 MHz}}\\ \cline{4-6}
         
         &&&  \multicolumn{2}{|c|}{\textbf{Data Rate}}& \multirow[c]{2}{*}[0in]{\shortstack{\textbf{Min.}\\ \textbf{SNR}}} \\ \cline{4-5}
       
       &&& \textbf{800ns}   & \textbf{400ns}&\\\hline
      
         $0$ &  BPSK & $1/2$  & $6.5$  & $7.2$ & $2$  \\\hline
         $1$ &  QPSK & $1/2$  & $13$  & $14.4$ & $5$  \\\hline
         $2$ &  QPSK & $3/4$  & $19.5$  & $21.7$ & $9$  \\\hline
         $3$ &  16-QAM & $1/2$  & $26$  & $28.9$ & $11$  \\\hline
         $4$ &  16-QAM & $3/4$  & $39$  & $43.3$ & $15$  \\\hline
         $5$ &  64-QAM & $2/3$  & $52$  & $57.8$ & $18$  \\\hline
         $6$ &  64-QAM & $3/4$  & $58.5$  & $65$ & $20$  \\\hline
         7 &  64-QAM & $5/6$  & $65$  & $72.2$ & $25$ \\\hline
    \end{tabular}
    \label{MCStable}
\end{table}

\textbf{Assumptions}: 1) We assume that the location of the {BS} (Alice), the MCS tables, the frequency of operation constitute public information, and thus, are known to Eve. 2) We consider a LOS scenario with one or more static UEs (Bob nodes). With this, we generate two sets of results: 3a) when Eve has the knowledge of the exact transmit powers of Alice and Bob, and 3b) when Eve knows the transmit powers of Alice and Bob in a narrow range\footnote{The exact transmit power of a BS in a cellular system is not known. That is, the BS downlink power is not fixed by 3GPP standards but rather dynamically allocated per user and resource block within hardware and regulatory limits, guided by principles in TS 36.213/38.213 and shaped in practice by factors such as CQI, {hybrid automatic repeat request (HARQ)}, fading, scheduling, and MCS, yet it is possible to reasonably infer the typical transmit power values of the BS and UE in a narrow range, by means of simple tools such as software-defined radio dongles.}. We believe the assumption 3(a), though not practical, serves as an upper bound on the performance of the proposed attack. 

\section{Modulation Classification \& MCS Table-based Passive Attack}
The proposed modulation classification-based passive attack consists of two distinct phases. 1) During downlink phase, the eavesdropper, Eve, intercepts the broadcast message from Alice to Bob, and does modulation classification on it in order to identify the modulation type of the transmitted signal. This allows Eve to utilize the MCS table to identify a small geographical region (basically, a ring) within the cell where presumably Bob is present. 2) During the uplink phase, Eve traverses through the geographical region narrowed down during the downlink phase. Eve continues its passive attack by intercepting the uplink signal. Eventually, Bob's location is obtained by Eve via a diverse set of strategies.

\subsection{Downlink Phase: Pseudo Ranging through Modulation Classification}

We first present sufficient details about the modulation classification method we have implemented, followed by the details of how Eve could map the detected modulation scheme to a ring, a process we call pseudo-ranging.

During the downlink phase, Alice sends a modulated signal to Bob using either phase-shift keying (PSK) or quadrature amplitude modulation (QAM). Specifically, Alice utilizes one of the specific modulation types from $4$ different modulation schemes, namely, BPSK, QPSK, 16-QAM, and 64-QAM (see Table~\ref{MCStable}). Eve sniffs the signal broadcast by Alice and carries out modulation classification. We note that modulation classification has been traditionally done through statistical methods which: i) first differentiate between the PSK and QAM using the fact that the PSK scheme leads to a constant-envelop signal, while the QAM scheme leads to a variable-envelop signal; ii) identify the order of modulation $M$ \cite{octaviapaper}. However, more recently, there is an increased interest in doing modulation classification through deep learning methods \cite{peng2021survey}. 
In this work, inline with recent research trends \cite{peng2021survey}, we implement a two-dimensional convolutional neural network (CNN) as the modulation classifier at Eve, which consists of four convolutional layers, three dropout layers, one average pooling layer, and two dense layers.


The input to the neural network is the in-phase and quadrature components of the sniffed signal. 
To train the CNN-based modulation classifier, an offline training dataset is constructed.
By moving Eve away from Alice in different directions, sufficient samples for each modulation type at varying SNR values are obtained. The final dataset contains $240,000$ samples for the four modulation types: BPSK, QPSK, 16-QAM, and 64-QAM over a range of SNR values from $2$~dB to $28$~dB. This is in accordance with the MCS Table in Table~\ref{MCStable}. The training dataset is balanced to mitigate bias by ensuring an equal number of samples ($60,000$) for each of the four classes. The training dataset is further split into training and validation datasets with an $80:20$ ratio. After offline training, the CNN-based modulation classifier at Eve is evaluated using an unbiased testing dataset of $10,000$ samples. The different scenarios considered, along with the classification results, are presented in Section~\ref{mod_results}.

\textit{Remark:} Due to the stealth nature of the attack considered in this work, Eve can simply acquire the required dataset offline and covertly without being exposed. Nevertheless, we acknowledge that in some scenarios it may be advantageous for Eve to adapt its CNN online. This can be achieved through: offline supervised pretraining followed by unsupervised domain adaptation or pseudo-labeling during deployment, transfer learning from publicly known modulation schemes defined in 3GPP standards using emulated signals, few-shot or continual learning to incrementally update the model when encountering new signal types or environments, and robustness-enhancing data augmentation during offline training to reduce the need for frequent retraining.

\begin{figure}
    \centering
    \includegraphics[width=0.9\linewidth]{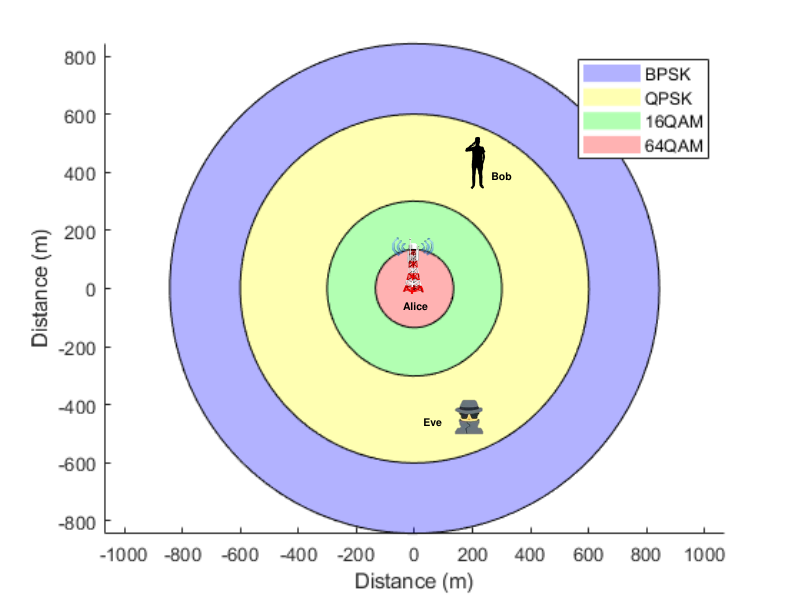}
    \caption{Possible rings of Bob based on the modulation type for $100$~GHz frequency and $400$~mW transmit power of Alice, within the coverage region of a single cell.}
    \label{fig:rings}
\end{figure}

Detection of a modulation classification through a CNN allows Eve to infer a small geographical region based on the possible {SNR} values that are associated with the classified modulation type in accordance with the MCS table given in Table~\ref{MCStable}. For example, when BPSK modulation scheme is detected, it implies that the operational SNR range of the downlink Alice-Bob channel is 2-5 dB (see Table~\ref{MCStable}). Eve plugs this SNR range into the Friis equation which returns upper and lower bounds on the distance between Alice and Bob. This corresponds to two concentric circles which together form a ring. Thus, the inferred regions of Bob are ring-shaped around Alice due to the LOS assumption and are depicted in Fig.~\ref{fig:rings} for a transmit frequency of $100$~GHz and transmit power of $400$~mW by Alice. We call this mapping process pseudo-ranging because it enables Eve to obtain loose upper and lower bounds on the distance between Alice and Bob. Pseudo-ranging greatly narrows down the search space for Eve from the full cell region down to a ring.


\begin{algorithm}[t!]
\caption{\textbf{:} UE localization during uplink phase}
\label{Algo2}
\textbf{Input:} BS/Alice's location, lower ($r_a$) and upper ($r_b$) limits of the location ring of UE/Bob, propagation model\\
\textbf{Initialize: } Step sizes $\Delta\theta_1$ and $\Delta\theta_2$, best and worst SNRs
\begin{algorithmic}[1]
\State Eve's initial location = {$\frac{r_b+r_a}{2}$}
\State $N_1\gets 360^{\circ}/\Delta\theta_1$
\State $i\gets 1$
\For{$i \leq N_1$}
\State {Compute  $SNR$ at Eve}
\If{$SNR >$  best $SNR$}
    \State {possible Bob's location $\gets$ current Eve's}
    \Statex {\hspace{1cm}location}
     \State {best $ SNR \gets$ current $SNR$}
\ElsIf{$SNR <$ best $SNR$}
    \State {\textbf{continue}}
\EndIf
\State $i=i+1$
\EndFor
\Statex \textbf{Output: } Initial estimate $(x_{B_c},y_{B_c})$ of Bob's location 

\State {Eve's location $\gets$ $(x_{B_c},y_{B_c})$}
\State {Compute  $SNR$ at Eve}
\State {Compute distance $d_{BE}$ between Bob \& Eve using Friis equation based pathloss model}
\State {Eve moves to a distance $d_{BE}$ from its current location}
\State $N_2\gets 360^{\circ}/\Delta\theta_2$
\For{$j=1:M$}
\State $i\gets 1$
\For{$i \leq N_2$}
\State {Compute  $SNR$ at Eve}
\If{$SNR >$  best $SNR$}
    \State {Possible Bob's location $\gets$ current Eve's}
    \Statex {\hspace{1.5cm}location}
     \State {best $ SNR \gets$ current $SNR$}
\ElsIf{$SNR <$ best $SNR$}
    \State {\textbf{continue}}
    
\EndIf
\State $i=i+1$
\EndFor
\State $j=j+1$
\EndFor
\end{algorithmic}
\textbf{Output: } Predicted location of Bob
\end{algorithm}

\subsection{Uplink Phase: UE Localization} \label{uplink_strategy}
After identifying the modulation scheme during the downlink phase, Eve moves to the ring where Bob is likely to be located and chooses a midpoint radius between the upper ($r_b$) and lower ($r_a$) limits of Bob's location ring, i.e., {$\frac{r_b+r_a}{2}$}. 
This can be easily realized by an eavesdropper UAV flying towards Bob's ring. Since Bob sends a signal to Alice at a significantly lower transmit power compared to the downlink transmission, this makes it more challenging to intercept the signal if Eve is far away from Bob. Thus, Eve wants to be in close proximity of Bob which helps it sniff a relatively high-quality copy of Bob's signal during the uplink phase. The sniffing of uplink channel helps Eve further narrow down the search space as follows. Inside the ring, Eve moves along a circular path of the midpoint radius. At each step, Eve sniffs the uplink signal of Bob, and computes the received SNR. After completion of the circular path, Eve declares its location where it receives the highest SNR as initial location $(x_{B_c},y_{B_c})$ of Bob.  




Further, as the SNR of Bob's signal is available at Eve, it again utilizes the Friis equation-based pathloss model to do coarse ranging, i.e., the measurement of approximate distance $d_{BE}$ between Eve and Bob. This allows Eve to move again in a smaller circle of origin $(x_{B_c},y_{B_c})$ and radius $d_{BE}$. This way, Eve declares the location where it receives the highest SNR as Bob's location. 
This step is repeated a few ($M$) times until the localization error is reduced sufficiently. 
This strategy considerably improves localization accuracy at the cost of additional resources required for distance $d_{BE}$ estimation. The detailed algorithm for the uplink phase is given in Algorithm~\ref{Algo2}. The entire process consists of refining/shrinking the search space, initially through modulation classification and subsequently utilizing the described strategy. 

\textit{Remark:} There exists a trade-off between the step size parameters $\Delta\theta_1$ and $\Delta\theta_2$ (see Algorithm 1) and the final localization accuracy achieved. That is, smaller step sizes yield higher accuracy, and vice versa. Further, smaller step sizes $\Delta\theta_2$ during the refinement stage are desirable. 

\begingroup
\setlength{\tabcolsep}{12pt} 
\renewcommand{\arraystretch}{1.5}
\begin{table*}[h]
    \centering
    \caption{Modulation classification accuracy for the different scenarios during downlink phase.}
    \begin{tabular}{|c|c|c|c|c|c|c|}
    \hline
         \multirow[c]3{*}[0in]{\textbf{Scenario}} & \multirow[c]{3}{*}[0in]{\textbf{Frequency (GHz)}}  & \multirow[c]{3}{*}[0in]{\textbf{Alice's Transmit Power (mW)}}   & \multicolumn{4}{|c|}{\textbf{Classification Accuracy (\%)}}\\ \cline{4-7}
      
       &&&  \multicolumn{3}{|c|}{\textbf{Testing Dataset}} & \multirow[c]{2}{*}[0in]{\textbf{Validation Dataset}} \\ \cline{4-6}
       
       &&&\textbf{Near Eve}   & \textbf{Mid Eve} & \textbf{Far Eve}&\\\hline
    
         a. & \multirow[c]{3}{*}[0in]{{5}} & {$200\pm10\%$} & {$87.12$} & {$76.44$}  & {$50.35$} & {$80.69$} \\ \cline{1-1}\cline{3-7}
        
         b. & & {$300\pm10\%$}   & {$93.37$} & {$76.98$} & {$48.45$} & {$83.41$} \\\cline{1-1}\cline{3-7}
      
         c. &  & {$400\pm10\%$}   & {$93.57$} & {$71.95$} & {$50.31$} & {$84.56$} \\\hline \hline
      
         d. & \multirow[c]{3}{*}[0in]{28} & {$200\pm10\%$}   & {$91.58$} & {$75.39$} & {$40.49$} & {$82.47$} \\ \cline{1-1}\cline{3-7}
        
         e. & & {$300\pm10\%$} & {$87.48$} & {$74.44$} & {$51.32$} & {$81.74$} \\\cline{1-1}\cline{3-7}
         f. &  & {$400\pm10\%$}  & {$85.25$} & {$71.35$} & {$52.51$} & {$80.46$} \\ \hline\hline
         g. & \multirow[c]{3}{*}[0in]{100} & {$200\pm10\%$}   & {$84.54$} & {$66.73$}  & {$41.88$} & {$81.44$} \\ \cline{1-1}\cline{3-7}
         h. & & {$300\pm10\%$} & {$95.37$} & {$74.44$} & {$49.98$} & {$84.49$}\\\cline{1-1}\cline{3-7}
         k. & & {$400\pm10\%$}  & {$81.85$} & {$74.68$} & {$47.26$} & {$82.42$} \\ \hline
         
    \end{tabular}
    \label{tab:results_phase1}
\end{table*}
\endgroup

\section{Performance Evaluation} \label{sim_results}

We evaluate the impact of the proposed passive attack across three different frequencies, i.e., 5 GHz (the microwave band), 28 GHz and 100 GHz (the millimeter-wave band). We also consider three different transmit powers of Alice, i.e., 200 mW, 300 mW, and 400 mW (with an uncertainty of 10\%). This results in nine distinct scenarios depicted in Table~\ref{tab:results_phase1}. The gains of transmit antennas at Alice, Bob, and receive antenna at Eve are set to 1. The transmit power of Bob is set to 0.5 mW (with an uncertainty of 10\%).

In the downlink phase, modulation classification results are obtained by testing the trained CNN on the corresponding testing dataset for each scenario. The CNN implementation and training are performed in JupyterLab, while the training and testing datasets are generated using MATLAB. For the uplink phase, the scenarios in Table~\ref{tab:results_phase1} are simulated using MATLAB Site Viewer, with the system model defined in Section~\ref{sys_model}. From the downlink phase, Eve is provided with the upper and lower limits of Bob's ring. Localization performance is assessed by repeating the localization task $1000$ times with Bob positioned arbitrarily each time. The following subsections provide key results and discussions for the different phases of the proposed approach.

\subsection{Downlink Modulation Classification Results} \label{mod_results}  
During the training of 2D CNN, the validation accuracy increased steadily up
to 10 epochs, beyond which signs of overfitting emerged. Therefore, training was limited to 10
epochs to ensure optimal generalization performance.
The testing dataset consists of intercepted downlink signals, and is generated for three distances of Eve from Alice (i.e., near, mid, and far), which helps capture the three different SNRs observed by Eve due to three different locations of Eve.

Table~\ref{tab:results_phase1} shows the classification accuracies of the CNN on the validation and testing datasets for the different considered frequency bands, and for three different transmit power of Alice (with 10\% uncertainty).
The validation accuracy exceeds $80\%$ for all the simulated scenarios. For the testing dataset, the accuracy of the model decreases significantly as Eve moves away from Alice, thereby reducing the SNR of the received signal at Eve. Moreover, with the increase in frequency, the total coverage area is reduced significantly owing to higher path loss at higher frequencies. 
Table~\ref{tab:results_phase1} illustrates that {\it Eve must stay close to Alice/BS in order to do high-quality pseudo-ranging while it intercepts the downlink signal}. This is because a near location of Eve results in accurate modulation classification, which in turn helps identify the correct location ring for Bob, and vice versa. 
Last but not the least, \textit{the uncertainty in the transmit power of Alice leads to a situation where different rings due to different modulation schemes used by Alice are overlapped}. However, this has almost no impact on the localization method detailed in Algorithm 1, due to the fact that the uplink phase is somewhat insensitive to the variations in the sizes of individual rings.

\subsection{Uplink Malicious Localization Results} \label{uplink_results}
\begin{figure}
  \centering
        
      
      
        
         
   \includegraphics[width=0.49\textwidth,height=0.9\textwidth,keepaspectratio]{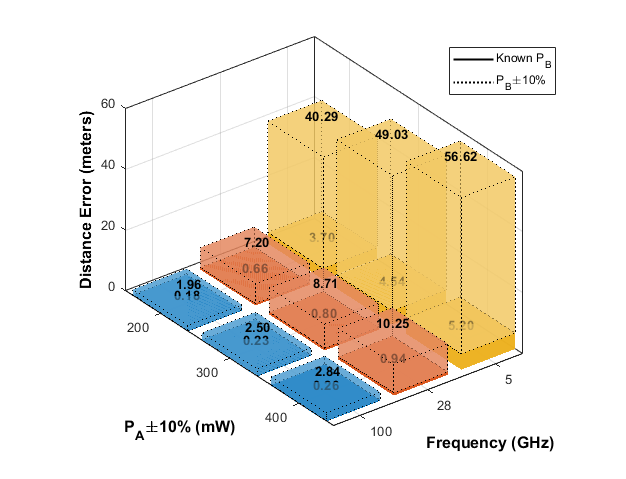} 

  \caption{Average distance error for different scenarios during the Uplink phase. $P_A$ and $P_B$ represent the powers of Alice and Bob, respectively.}
  \label{fig:strat2result}
\end{figure}

To evaluate the performance of the proposed malicious localization attack during the uplink phase, we conduct Monte-Carlo simulations with $1000$ iterations. In each iteration, Eve utilizes the upper and lower limits of Bob's ring obtained via modulation classification and follows the strategy proposed in Algorithm~\ref{Algo2} in order to localize Bob, who is randomly positioned within the ring. The localization performance is measured by calculating the average distance error between Eve's predicted locations and Bob's actual locations over the $1000$ runs. The results of the Monte-Carlo simulations are summarized in Fig.~\ref{fig:strat2result}. 
It can be seen that \textit{the distance error increases when the transmit power of Alice increases} from ($200\pm10\%$)~mW to ($400\pm10\%$)~mW. This is mainly due to the fact that wireless signals can propagate to further distances with higher transmit powers, hence expanding the coverage region. Thus, the ring in which Bob potentially exists gets bigger, thereby expanding the search space for Eve and increasing; as a result, the localization error. 
Moreover, the distance error decreases with the increase in frequency from microwave ($5$~GHz) to mmWave ($28$~GHz and $100$~GHz) frequencies. 
The location ring reduces at mmWave frequencies due to higher attenuation experienced at high frequencies. Thus, \textit{the higher (mm-wave) frequencies are more prone to such threats}. 
Last but not the least, Fig. \ref{fig:strat2result} reveals that the localization error increases by many folds when Eve knows the transmit power of Bob in a range (with 10\% uncertainty), compared to the case when Eve knows the exact transmit power of Bob. Thus, we conclude that \textit{the proposed attack leads to fine-grained localization of Bob when transmit power of Bob is known exactly, and coarse-level proximity detection of Bob when Bob's transmit power is known within a range}.


\begin{figure*}[h]
  \centering
  \subfloat[Error for Bob1 (two Bobs case).]{\includegraphics[width=0.33\textwidth,height=0.8\textwidth,keepaspectratio]{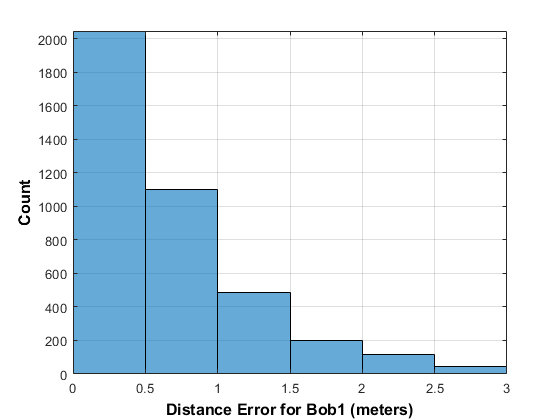}\label{bob1}}
  \subfloat[Error for Bob2 (two Bobs case).]{\includegraphics[width=0.33\textwidth,height=0.8\textwidth,keepaspectratio]{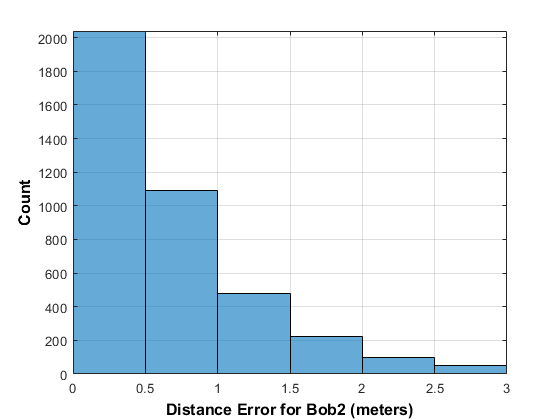}\label{bob2}}
  \subfloat[Distance error for scenario (d) when Eve employs a 10-element ULA.]{\includegraphics[width=0.33\textwidth,height=0.8\textwidth,keepaspectratio]{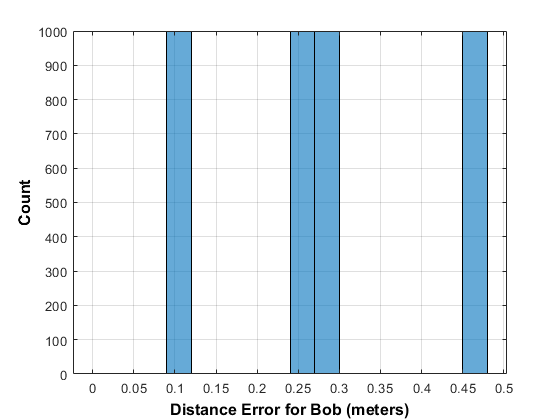}\label{fig: ULA}}
    \caption{Histograms of distance error for the case of two Bobs (scenario (d)) and with Eve having multiple antennas. 
    }
\end{figure*}


\subsection{Localization of Multiple UEs} \label{multibob}

To evaluate the generalization performance of the proposed attack, we consider a multi-user scenario where Eve attempts to localize multiple UEs/Bobs. This setup reflects a more realistic environment, such as a cellular or WiFi network, where several users connect to the same base station or access point. In the uplink phase, Eve would require increased receiver bandwidth proportional to the number of Bobs being monitored and localized. Additionally, this scenario demands greater resources, including a larger number of Eves, longer time requirements, and increased processing power. For example, if two Bobs are located in different rings, either a minimum of two Eves (one in each ring) is needed, or a single Eve could localize them sequentially, though this would be more time-intensive.

In simulation, Eve aims to localize two users, referred to as Bob1 and Bob2, who are communicating with Alice/BS using different orthogonal frequency-division multiple access (OFDMA) sub-carriers. Since the sub-carrier spacing is relatively small in 5G systems (on the order of kHz), the possible location rings for Bob1 and Bob2, determined during the modulation classification step in the downlink phase, remain unchanged. In the uplink phase, where both Bobs are to be localized within their respective rings, we employ Algorithm~\ref{Algo2}, similar to the single-user case. The key difference is that Eve now intercept two UE channels to take two SNR measurements at each step instead of one.

To evaluate the effectiveness of the proposed attack to localize two UEs, we chose scenario (d) from Table~\ref{tab:results_phase1}, i.e., we set the center frequency to $28$~GHz and transmit power of Alice to $200$~mW. 
We ran Monte Carlo simulations, repeating the experiment $1000$ times. In each iteration, Eve attempted to localize two Bobs, each randomly positioned within one of the several rings. The average localization errors are $0.66$~m for Bob1 and $0.67$~m for Bob2, which are comparable to the results obtained for a single Bob for the same scenario (see Fig.~\ref{bob1} and Fig~\ref{bob2}). Thus, the proposed passive attack could localize multiple UEs without any performance degradation, but at the expense of more compute resources at Eve.


\subsection{Eve with Multiple Antennas} \label{mimoEve}

We now evaluate the situation whereby Eve employs a uniform linear array (ULA) and performs phased array processing in order to determine the direction of arrival (DoA) of Bob's signal during uplink phase. 
When equipped with a ULA, Eve first obtains an initial estimate of location of Bob using Algorithm~\ref{Algo2}. Then, Eve utilizes phased array processing method in order to obtain a DoA estimate $\theta$ from Bob's signal. This DoA estimate $\theta$ coupled with the distance estimate $R$ is utilized by Eve to move by a distance $R$ directly towards Bob's direction $\theta$ in a single-shot.

To evaluate the efficacy of the proposed attack by an Eve with multiple antennas, we chose scenario (d) from Table~\ref{tab:results_phase1}, i.e., we set the center frequency to $28$~GHz and transmit power of Alice to $200$~mW. Eve was equipped with a $10$-element ULA, with element spacing set to half the wavelength. 
We note that though a number of algorithms such as beam scan, multiple signal classification (MUSIC), minimum variance distortionless response (aka CAPON), and amplitude and phase estimation (APES) exist, we chose the root-MUSIC algorithm to estimate the DoA, due to its ability to resolve closely spaced signals and perform well at low SNRs. 

We conducted Monte-Carlo simulations, with each experiment repeated $1000$ times. 
In each experiment, Eve obtained the upper and lower bounds on Bob's location ring, through modulation classification during the downlink phase. Eve then applied Algorithm~\ref{Algo2} to compute Bob's intermediate position within the ring. This was followed by the estimation of DoA $\theta$ and distance $R$ to accurately localize Bob at location $Re^{j\theta}$. As shown in Fig.~\ref{fig: ULA}, the localization error is significantly reduced compared to the single-antenna case under similar conditions. Specifically, we achieve a much lower average localization error of $0.27$~m compared to $0.66$~m for the single-antenna case in scenario (d), as illustrated in Fig.~\ref{fig:strat2result}. This demonstrates that employing multiple antennas at the adversarial node Eve enhances the performance of the proposed attack, improving its ability to extract location information.

\subsection{Computational complexity/latency of the proposed attack}
The computational complexity of the proposed passive sensing attack is determined by two main components: (i) modulation classification during the downlink phase and (ii) user localization during the uplink phase. For the downlink phase, complexity arises from the custom CNN used for modulation classification. Summing across all layers (4 Conv2D, 1 AveragePooling, Flatten, and 2 Dense layers), the total CNN complexity is $59.06$ MFLOPs for an input size of \texttt{(2, 128, 1)}. 
In the uplink phase, the localization procedure (Algorithm~1) has complexity $O\big(\tfrac{1}{\Delta\theta_1} + M \cdot \tfrac{1}{\Delta\theta_2}\big)$, due to an initial coarse angular search followed by refinement. With $M=O(1)$ iterations during refinement step, this reduces to $O\big(\tfrac{1}{\Delta\theta_1} + \tfrac{1}{\Delta\theta_2}\big)$; for instance, setting $\Delta\theta_1=5^\circ$ and $\Delta\theta_2=1^\circ$ yields tractable overhead. Combining both phases, the total complexity is modest, requiring only tens of MFLOPs. If Eve utilizes a GPU capable of $10^{12}$ FLOPs/sec, then execution time is as little as a few milliseconds, suggesting feasibility of the proposed attack for edge-device deployment. Parallelization or hardware acceleration can further improve the performance.

\section{Discussion}
\subsection{Scenario of non-line-of-sight Users}

In many real-world environments--particularly dense urban and complex indoor settings--{non-line-of-sight (NLOS)} propagation dominates. For the NLOS scenario, the mapping of a modulation scheme to a nicely shaped geographical region, i.e., a ring, does not hold anymore. However, we believe that Eve could still do malicious proximity detection (if not fine-grained localization) under the NLOS scenario using radio fingerprinting or statistical path-loss models such as log-normal shadowing. We could foresee a three-phase process: (i) offline radio map construction, where high-resolution NLOS modulation maps are generated through measurement campaigns or electromagnetic ray-tracing (e.g., NVIDIA Sionna with 3D urban geometry from OpenStreetMap), capturing irregular multipath pockets and shadow zones; (ii) candidate region identification, where intercepted downlink signals are classified for the modulation scheme in use and matched against the NLOS map; and (iii) precision refinement, where Eve deploys multiple static or mobile eavesdropper nodes in the identified region according to a fixed or random geometry, each with multiple antennas to infer rich spatio-temporal insights from uplink signals, and utilizes machine learning models to obtain a reasonable estimate of the locations of the UEs. Nevertheless, the accuracy for NLOS scenario is likely to remain highly dependent on environment-specific calibration.



\subsection{Scenario of LOS Mobile Users}
In case of LOS mobile UEs such as cars, UAVs, pedestrians on the road, a hybrid Markov chain plus Kalman filtering based framework could be employed by Eve for malicious tracking of mobile UEs. Specifically, for a given UE, the Markov chain framework will keep track of the state of the UE (i.e., in which ring the UE lies), while the Kalman filter will do mobility tracking of the UE (i.e., speed and direction of heading).

\subsection{Synthesis of Malicious Digital Twin Map}

It is possible to create a digital twin of the environment by mere passive sensing only, i.e., by combining our proposed localization method with a tracking mechanism like Bayesian filtering (e.g., Kalman filtering) along with detailed 3D mapping. This digital replica enables real-time tracking of user movements, and therefore, has both positive and negative implications in a number of scenarios. For example, the mobile operators could utilize the digital twin to enhance the efficiency of radio resource management by enabling the dynamic allocation of resources to areas with higher user density. Further, the mobile operators could also utilize the digital twin to realize additional revenue streams by providing insights to third party businesses about the mobility patterns of people in a region. On the other hand, governments could utilize the digital twin approach for monitoring various kinds of aggregate level behaviors of people in a neighborhood. Need not to say that the adversaries could also synthesize and maintain a digital twin in a fully passive and covert manner, and utilize it in a number of unforeseen malicious ways to cause various kinds of security threats, disrupt the social fabric, inflict financial losses, and more.  


\section{Conclusion} \label{sec:conclusion}

This paper argues that the openness of wireless cellular (3GPP Releases 16-20) and WLAN (IEEE 802.11xx) standards could make them vulnerable to various kinds of malicious attacks by adversaries. Specifically, this work demonstrates how an eavesdropper can passively localize a stationary LOS user in a cellular or WiFi network through a two-phase attack. By intercepting downlink communication and utilizing modulation classification, the eavesdropper can estimate the user's location within a ring. In the second phase, more precise localization is achieved by analyzing the user's uplink data. Our simulations validate the effectiveness of this attack in single-user, multi-user, and multiple-antenna scenarios. While this attack currently applies to LOS conditions with stationary UEs, future research could extend it to NLOS scenarios and mobile UEs (e.g., cars, UAVs, pedestrians), posing broader security risks to wireless networks. While private 5G networks are immune to this attack, public 5G/WiFi networks are not.


\footnotesize{
\bibliographystyle{IEEEtran}
\bibliography{references}
}
\vspace{-3em}
\begin{IEEEbiographynophoto}{Ali Hanif}
received his B.E. and M.S. degrees in Avionics Engineering from the National University of Sciences and Technology (NUST), Pakistan, in 2014 and 2021, respectively. Currently, he is a Ph.D. student with the Information Science Lab at King Abdullah University of Science and Technology (KAUST). His research interests include wireless communications, radar systems, and deep learning.
\end{IEEEbiographynophoto}
\vspace{-5em}
\begin{IEEEbiographynophoto}{Abdulrahman Katranji}
received his B.Sc. degree in electrical engineering from King Fahad university of petroleum and minerals, KSA in 2024. He is currently an MS student with the same university.
\end{IEEEbiographynophoto}
\vspace{-5em}
\begin{IEEEbiographynophoto}{Nour Kouzayha}
received her Ph.D. in electrical and computer engineering from the American University of Beirut, Lebanon, in 2018. She is currently a research scientist at KAUST. Her research interests are in the broad area of wireless communications and networking with a special focus on 5G/6G networks, the Internet of Things, THz communications, and non-terrestrial networks.
\end{IEEEbiographynophoto}
\vspace{-5em}
\begin{IEEEbiographynophoto}{Muhammad  Mahboob Ur  Rahman}
received his Ph.D. degree in Electrical and Computer engineering from The University of Iowa, Iowa, IA, USA in 2013. In summer 2013, he was a Research Intern with the Wireless Systems Lab, Nokia Research Centre, Berkeley, CA, USA. He then joined the Communication Theory Laboratory, KTH, Stockholm, Sweden, as a Post-Doctoral researcher. From 2016 to 2022, he worked as Assistant Professor with the Electrical Engineering Department, Information Technology University, Lahore, where he led the Wireless Solutions (WiSo) Research Laboratory. He is currently a research scientist at KAUST, KSA. His research interests include: physical layer security, AI for 6G, quality-of-service analysis, and open RAN.
\end{IEEEbiographynophoto}
\vspace{-5em}
\begin{IEEEbiographynophoto}{Tareq Y. Al-Naffouri}
received the B.S. degrees in mathematics and electrical engineering (with first honors) from King Fahd University of Petroleum and Minerals, Saudi Arabia, the M.S. degree in electrical engineering from the Georgia Institute of Technology, and the Ph.D. degree in electrical engineering from Stanford University, Stanford in 2004. He was a visiting scholar at California Institute of Technology, Pasadena, CA in 2005 and summer 2006. He was a Fulbright scholar at the University of Southern California in 2008. He is currently a Professor at the Electrical Engineering Department, KAUST. His research interests lie in the areas of sparse, adaptive, and statistical inference/learning and their applications to wireless communications, localization, smart cities, and smart health. He has over 370 publications in journal and conference proceedings and 24 issued/pending patents. He has won the IEEE Education Society Chapter Achievement Award (2008), Almarie Award for Innovative research in communication (2009), and Abdul Hameed Shoman Prize for innovative research in IoT (2022).
\end{IEEEbiographynophoto}


\end{document}